\documentclass[11pt]{article}
\usepackage{times}
\usepackage{eepic,epic}
\usepackage{latexsym}
\usepackage{amsfonts}
\usepackage{amsmath}
\usepackage{amssymb}
\usepackage{amsopn}
\usepackage{graphicx}
\makeatletter%
\def\nottoobig#1{{\hbox{$\left#1\vcenter to1.111\ht\strutbox{}\right.\n@space$}}}
\makeatother%

\makeatother%

\makeatletter%

\newcount\hour  \newcount\minutes  \hour=\time  \divide\hour by 60
\minutes=\hour  \multiply\minutes by -60  \advance\minutes by \time
\def\mmmddyyyy{\ifcase\month\or Jan\or Feb\or Mar\or Apr\or May\or Jun\or Jul\or
  Aug\or Sep\or Oct\or Nov\or Dec\fi \space\number\day, \number\year}
\def\hhmm{\ifnum\hour<10 0\fi\number\hour :%
  \ifnum\minutes<10 0\fi\number\minutes}
\def\Draft{{\it Draft of \mmmddyyyy}}

\def\ps@jtsheadings{%
\def\@oddhead{\it\rightmark\hfil\rm\thepage}%
\def\@oddfoot{\hfil\Draft}%
\if@twoside%
\def\@evenhead{\rm\thepage\hfil\it\leftmark}%
\def\@evenfoot{\Draft\hfil}%
\else
\let\@evenhead\@oddhead%
\let\@evenfoot\@oddfoot%
\fi%
}
\def\ps@jtsplain{%
\def\@oddhead{\hfil\Draft}%
\def\@oddfoot{\hfil\rm\thepage\hfil}%
\let\@evenfoot\@oddfoot%
\if@twoside \def\@evenhead{\Draft\hfil} \else \let\@evenhead\@oddhead \fi
}

\def\chaptermark#1{\markboth{\thechapter.\ #1}{\thechapter.\ #1}}%
\def\sectionmark#1{\markright{\thesection.\ #1}}

\def\section{\@startsection {section}{1}{\z@}
    {3.5ex plus1ex minus.2ex}{2.3ex plus.2ex}{\Large\bf}}
\def\subsection{\@startsection{subsection}{2}{\z@}
    {3.25ex plus1ex minus.2ex}{1.5ex plus.2ex}{\large\bf}}
\def\subsubsection{\@startsection{subsubsection}{3}{\z@}
    {3.25ex plus1ex minus.2ex}{1.5ex plus.2ex}{\normalsize\bf}}
\def\paragraph{\@startsection{paragraph}{4}{\z@}
    {3.25ex plus1ex minus.2ex}{1em}{\normalsize\bf}}
\def\subparagraph{\@startsection{subparagraph}{4}{\parindent}
    {3.25ex plus1ex minus.2ex}{1em}{\normalsize\bf}}

\makeatother%

\makeatletter \@beginparpenalty=10000 \makeatother

\def\underl#1 {\leavevmode\let\first=\relax\underli #1 }
\def\underli#1 {\ifx&#1\let\next=\relax\unskip
                \else\let\next=\underli\first\ulinebox{#1}\fi\let\first=\undersp\next}
\def\undersp{\penalty50\ulinebox{\space}\penalty50}
\def\ulinebox#1{\vtop{\hbox{\strut#1}\hrule}}%
\def\unice#1 {\underl #1 & }
\def\desclabel#1{\bf #1\hfil}
\def\desc{\list{}{%
\labelwidth=\leftmargin
\advance \labelwidth by -\labelsep
\let \makelabel=\desclabel}}

\makeatletter %

\newlength{\filength}
\newsavebox{\gcbox}
\sbox{\gcbox}{\framebox[\filength]{\rule{0ex}{2ex}}}

\newlength{\leftjustindent}
\newlength{\@leftjustindent}
\def\leftjust{\let\\\@leftjustcr\let\end\@endleftjust
  \addtolength{\@leftjustindent}{\leftjustindent}
  \vcenter\bgroup
  \halign\bgroup
    \hbox to\displaywidth{
      \rule{\@leftjustindent}{0ex}$\displaystyle##$\hfill
      }\crcr
}
\def\endleftjust{\crcr\egroup\egroup\endgroup}
\def\@endleftjust#1{\crcr\egroup\egroup\@checkend{#1}\endgroup}
\def\@leftjustcr{\crcr}

\newenvironment{proof}{\noindent{\bf Proof.}\hspace*{1em}}{\qed\bigskip}
\newenvironment{proofoflemma}[1]{\noindent{\bf Proof of Lemma~#1.}\hspace*{1em}}{\medskip}
\newtheorem{theorem}{Theorem}[section]

\newcommand{\qedblob}{\mbox{\rule[-1.5pt]{5pt}{10.5pt}}}
\def\literalqed{{\ \nolinebreak\hfill\mbox{\qedblob\quad}}}

\def\qed{\literalqed}
\newcommand{\qedoflemma}[1]{{\ \nolinebreak\hfill\mbox{\qedblob~$_{\mbox{\scriptsize{}Lemma~#1}}$\quad}}}

\newtheorem{lemma}[theorem]{Lemma}

\newcommand{\singlespacing}{\let\CS=
\@currsize\renewcommand{\baselinestretch}{1}\tiny\CS}
\newcommand{\singlespacingplus}{\let\CS=
\@currsize\renewcommand{\baselinestretch}{1.25}\tiny\CS}
\newcommand{\doublespacing}{\let\CS=
\@currsize\renewcommand{\baselinestretch}{1.75}\tiny\CS}
\newcommand{\draftspacing}{\let\CS=
\@currsize\renewcommand{\baselinestretch}{2.0}\tiny\CS}
\newcommand{\foospacing}{\let\CS=
\@currsize\renewcommand{\baselinestretch}{1.05}\tiny\CS}

\makeatother%

\mathcode`\0="0030      %
\mathcode`\1="0031
\mathcode`\2="0032
\mathcode`\3="0033
\mathcode`\4="0034
\mathcode`\5="0035
\mathcode`\6="0036
\mathcode`\7="0037
\mathcode`\8="0038
\mathcode`\9="0039

\newtheorem{definition}[theorem]{Definition}

\makeatletter
\clubpenalty=\@highpenalty
\widowpenalty=\@highpenalty
\makeatother

\makeatletter
\newcommand{\niceonespacing}{\let\CS=\@currsize\renewcommand{\baselinestretch}{1.1}\tiny\CS}\newcommand{\nicetwospacing}{\let\CS=\@currsize\renewcommand{\baselinestretch}{1.2}\tiny\CS}
\newcommand{\nicethreespacing}{\let\CS=\@currsize\renewcommand{\baselinestretch}{1.3}\tiny\CS}
\newcommand{\singlespacingplusplus}{\let\CS=\@currsize\renewcommand{\baselinestretch}{1.35}\tiny\CS}
\newcommand{\nicefourspacing}{\let\CS=\@currsize\renewcommand{\baselinestretch}{1.4}\tiny\CS}
\newcommand{\nicefivespacing}{\let\CS=\@currsize\renewcommand{\baselinestretch}{1.5}\tiny\CS}
\newcommand{\nicesixpacing}{\let\CS=\@currsize\renewcommand{\baselinestretch}{1.6}\tiny\CS}
\makeatother

\makeatletter
\def\@cite#1#2{[#1\if@tempswa , #2\fi]}
\makeatother

\makeatletter
\def\@citex[#1]#2{\if@filesw\immediate\write\@auxout{\string\citation{#2}}\fi
  \def\@citea{}\@cite{\@for\@citeb:=#2\do
    {\@citea\def\@citea{,\linebreak[0]}\@ifundefined
       {b@\@citeb}{{\bf ?}\@warning
       {Citation `\@citeb' on page \thepage \space undefined}}%
\hbox{\csname b@\@citeb\endcsname}}}{#1}}
\makeatother

\makeatletter
\def\ps@thesis{\def\@oddhead{\hfil\rm\thepage\hfil}\def\@oddfoot{}\def\@evenhead{\hfil\rm\thepage\hfil}\def\@evenfoot{}\def\chaptermark##1{}\def\sectionmark##1{}}
\makeatother

\makeatletter
\def\foobarpt{\textfont\z@\tenrm 
  \scriptfont\z@\ninrm \scriptscriptfont\z@\sevrm
\textfont\@ne\tenmi \scriptfont\@ne\ninmi \scriptscriptfont\@ne\sevmi
\textfont\tw@\tensy \scriptfont\tw@\ninsy \scriptscriptfont\tw@\sevsy
\textfont\thr@@\tenex \scriptfont\thr@@\tenex \scriptscriptfont\thr@@\tenex
\def\unboldmath{\everymath{}\everydisplay{}\@nomath\unboldmath
          \textfont\@ne\tenmi 
          \textfont\tw@\tensy \textfont\lyfam\tenly
          \@boldfalse}\@boldfalse
\def\boldmath{\@ifundefined{tenmib}{\global\font\tenmib\@mbi\@magscale1\global
        \font\tensyb\@mbsy \@magscale1\global\font
         \tenlyb\@lasyb\@magscale1\relax\@addfontinfo\@xiipt
              {\def\boldmath{\everymath
                {\mit}\everydisplay{\mit}\@prtct\@nomathbold
                \textfont\@ne\tenmib \textfont\tw@\tensyb 
                \textfont\lyfam\tenlyb\@prtct\@boldtrue}}}{}\@xiipt\boldmath}%
\def\prm{\fam\z@\tenrm}%
\def\pit{\fam\itfam\tenit}\textfont\itfam\tenit \scriptfont\itfam\ninit
   \scriptscriptfont\itfam\sevit
\def\psl{\fam\slfam\tensl}\textfont\slfam\tensl 
     \scriptfont\slfam\tensl \scriptscriptfont\slfam\tensl
\def\pbf{\fam\bffam\tenbf}\textfont\bffam\tenbf 
   \scriptfont\bffam\ninbf \scriptscriptfont\bffam\ninbf 
\def\ptt{\fam\ttfam\tentt}\textfont\ttfam\tentt
   \scriptfont\ttfam\nintt \scriptscriptfont\ttfam\nintt 
\def\psf{\fam\sffam\tensf}\textfont\sffam\tensf
    \scriptfont\sffam\tensf \scriptscriptfont\sffam\tensf
\def\psc{\@getfont\psc\scfam\@xiipt{\@mcsc\@magscale1}}%
\def\ly{\fam\lyfam\tenly}\textfont\lyfam\tenly 
   \scriptfont\lyfam\ninly \scriptscriptfont\lyfam\sevly
 \@setstrut \rm}

\makeatother

\newcommand{\p}{\mbox{\rm P}}

\newcommand{\np}{\mbox{\rm NP}}

\newcommand{\scriptnp}{\mbox{\scriptsize\rm NP}}

\def\pair#1{{{\langle\!\!~#1~\!\!\rangle}}}

\newcommand{\parallelnp}{\mbox{$\p_{\|}^{\scriptnp}$}}

\newcommand\seq{\subseteq}

\newcommand{\DodgsonScore}{\mbox{\rm DodgsonScore}}
\newcommand{\YoungScore}{\mbox{\rm YoungScore}}

\newenvironment{block}{\begin{list}{\hbox{}}{\leftmargin 1em
    \itemindent -1em \topsep 0pt \itemsep 0pt \partopsep 0pt}}{\end{list}}

\makeatletter
\def\@listI{\leftmargin\leftmargini \parsep 4.5pt plus 1pt minus 1pt\topsep
6pt plus 2pt minus 2pt \itemsep  2pt plus 2pt minus 1pt}

\let\@listi\@listI
\@listi
\makeatother

\begin{document}

\title{Exact Complexity 
of the Winner Problem for Young Elections\thanks{Supported in part by grant
    NSF-INT-9815095/DAAD-315-PPP-g\"u-ab.}}

\author{
  J\"{o}rg Rothe\thanks{Corresponding author.  
  Email: ${\tt rothe@cs.uni\mbox{-}duesseldorf.de}$.}
  \hspace*{4ex}
  Holger Spakowski\thanks{Email: 
  ${\tt spakowsk@cs.uni\mbox{-}duesseldorf.de}$.} \\
  Abteilung f\"ur Informatik \\
  Heinrich-Heine-Universit\"at D\"usseldorf \\
  40225 D\"usseldorf, Germany
  \and
  J\"{o}rg Vogel\thanks{Email:
  ${\tt vogel@minet.uni\mbox{-}jena.de}$.} \\
  Institut f\"ur Informatik \\
  Friedrich-Schiller-Universit\"at Jena\\
  07740 Jena, Germany
}

\date{December 20, 2001}

\maketitle

\begin{abstract}
  In 1977, Young proposed a voting scheme that extends the Condorcet Principle
  based on the fewest possible number of voters whose removal yields a
  Condorcet winner.  We prove that both the winner and the ranking problem for
  Young elections is complete for~$\parallelnp$, the class of problems
  solvable in polynomial time by parallel access to~$\np$.  Analogous
  results for Lewis Carroll's 1876 voting
  scheme were recently established by Hemaspaandra et al.  In contrast, we
  prove that the winner and ranking problems in Fishburn's homogeneous variant
  of Carroll's voting scheme can be solved efficiently by linear programming.
\end{abstract}

\section{Introduction}

More than a decade ago, Bartholdi, Tovey, and
Trick~\cite{bar-tov-tri:j:who-won,bar-tov-tri:j:manipulating,bar-tov-tri:j:control}
initiated the study of electoral systems with respect to their computational
properties.  In particular, they proved NP hardness lower 
bounds~\cite{bar-tov-tri:j:who-won} for
determining the winner in the voting schemes proposed by Dodgson (more
commonly known by his pen name, Lewis Carroll) and by
Kemeny, and they studied complexity issues
related to the problem of manipulating elections by strategic
voting~\cite{bar-tov-tri:j:manipulating,bar-tov-tri:j:control}.
Since then, a number of related results and improvements of their results have
been obtained.  Hemaspaandra, Hemaspaandra, and
Rothe~\cite{hem-hem-rot:j:dodgson} classified both the winner and
the ranking problem for Dodgson elections by proving them complete
for~$\parallelnp$, the class of problems solvable in polynomial time by
parallel access to an NP oracle.  E.  Hemaspaandra (as cited
in~\cite{hem-hem:c:electoral-systems-survey}) and Spakowski and
Vogel~\cite{spa-vog:c:kemeny} obtained the analogous result
for Kemeny elections; a joint paper by E. Hemaspaandra, Spakowski, and
Vogel is in preparation.  
For many further interesting results and the state of the
art regarding computational politics, we
refer to the survey~\cite{hem-hem:c:electoral-systems-survey}.

In this paper, we study complexity issues related to Young and Dodgson
elections.  In 1977, Young proposed a voting scheme that extends the Condorcet
Principle based on the fewest possible number of voters whose removal makes a
given candidate $c$ the Condorcet winner, i.e., $c$ defeats all other
candidates by a strict majority of the votes.  We prove that both the
winner and the ranking problem for Young elections is complete
for~$\parallelnp$.  To this end, we give a reduction from the problem ${\tt
  Maximum}$ ${\tt Set}$ ${\tt Packing}$ ${\tt Compare}$, which we also prove
$\parallelnp$-complete.  

In Section~\ref{sec:homogeneous}, we study a homogeneous variant of
Dodgson elections that was introduced by Fishburn~\cite{fis:j:condorcet}.  In
contrast to the above-mentioned result of Hemaspaandra et
al.~\cite{hem-hem-rot:j:dodgson}, we show that both the winner and the ranking
problem for Fishburn's homogeneous Dodgson elections can be solved efficiently 
by a linear program that is based on an integer linear program of Bartholdi et
al.~\cite{bar-tov-tri:j:who-won}.

\section{Complexity of the Winner Problem for Young Elections} 
\label{sec:votingschemes}

\subsection{Some Background from Social Choice Theory}
\label{sec:social-choice}

Let $C$ be the set of all candidates (or alternatives).  We assume that each
voter has strict preferences over the candidates.  Formally, the preference
order of each voter is strict (i.e., irreflexive and antisymmetric),
transitive, and complete (i.e., all candidates are ranked by each voter).  An
election is given by a {\em preference profile}, a pair $\pair{C,V}$ such that
$C$ is a set of candidates and $V$ is the multiset of the voters' preference
orders on~$C$.  Note that distinct voters may have the same preferences over
the candidates.  A {\em voting scheme\/} (or {\em social choice function}, SCF
for short) is a rule for how to determine the winner(s) of an election; i.e.,
an SCF maps any given preference profile to society's aggregate {\em choice
  set}, the set of candidates who have won the election.  For any SCF $f$ and
any preference profile~$\pair{C,V}$, $f(\pair{C,V})$ denotes the set of
winning candidates.  For example, an election is won according to the {\em
  majority rule\/} by any candidate who is preferred over any other candidate
by a strict majority of the voters.  Such a candidate is called the Condorcet
winner.

In 1785, Marie-Jean-Antoine-Nicolas de Caritat, the Marquis de Condorcet,
noted in his seminal essay~\cite{con:b:condorcet-paradox} that whenever there
are at least three candidates, say~$A$, $B$, and~$C$, the majority rule may
yield cycles: $A$ defeats~$B$ and $B$ defeats~$C$, and yet $C$ defeats~$A$.
Thus, even though each individual voter has a rational (i.e., transitive or
non-cyclic) preference order, society may behave irrationally and Condorcet
winners do not always exist.  This observation is known as the Condorcet
Paradox.  The {\em Condorcet Principle\/} says that for each preference
profile, the winner of the election is to be determined by the majority rule.
An SCF is said to be a {\em Condorcet SCF\/} if and only if it respects the
Condorcet Principle in the sense that the Condorcet winner is elected 
whenever he or she exists.  
Note that Condorcet winners are uniquely determined if they exist.

Many Condorcet SCFs have been proposed in the social choice literature; for an
overview of the most central ones, we refer to the work of
Fishburn~\cite{fis:j:condorcet}.  They extend the Condorcet Principle in a way
that avoids the troubling feature of the majority rule.  In this paper, we
will focus on only two such Condorcet SCFs, the Dodgson voting
scheme~\cite{dod:unpubMAYBE:dodgson-voting-system} and the Young voting
scheme~\cite{you:j:extending-condorcet}.

In 1876, Charles L. Dodgson
(better known by his pen name, Lewis Carroll) proposed a voting 
scheme~\cite{dod:unpubMAYBE:dodgson-voting-system} that
suggests that we remain most faithful to the Condorcet Principle if the
election is won by any candidate who is ``closest'' to being a Condorcet
winner.  To define ``closeness,'' each candidate $c$ in a given election
$\pair{C, V}$ is assigned a score, denoted $\DodgsonScore(C,c,V)$, which is
the smallest number of sequential interchanges of adjacent candidates in the
voters' preferences that are needed to make $c$ a Condorcet winner.  Here, one
interchange means that in (any) one of the voters two adjacent candidates are
switched.  A {\em Dodgson winner\/} is any candidate with minimum Dodgson
score.  Using Dodgson scores, one can also tell who of two given candidates is
ranked better according to the Dodgson SCF.

Young's approach to extending the Condorcet Principle is reminiscent of
Dodgson's approach in that it is also based on altered profiles.  Unlike
Dogson, however, Young~\cite{you:j:extending-condorcet} suggests that we
remain most faithful to the Condorcet Principle if the election is won by any
candidate who is made a Condorcet winner by {\em removing the fewest possible
  number of voters}, instead of doing the fewest possible number of switches
in the voters' preferences.  For each candidate $c$ in a given preference
profile~$\pair{C,V}$, define $\YoungScore(C,c,V)$ to be the size of a
largest subset of $V$ for which $c$ is a Condorcet winner.  A {\em Young
  winner\/} is any candidate with a maximum Young score.

Homogeneous variants of these voting schemes will be defined in
Section~\ref{sec:homogeneous}.

\subsection{Complexity Issues Related to Voting Schemes}
\label{sec:issues}

To study computational complexity issues related to Dodgson's voting scheme,
Bartholdi, Tovey, and Trick~\cite{bar-tov-tri:j:who-won} defined the following
decision problems.

\begin{quote}
  ${\tt Dodgson}$ ${\tt Winner}$ \\[1ex]
  {\bf Instance:} A preference profile $\pair{C,V}$ and
  a designated candidate~$c \in C$. \\
  {\bf Question:} Is $c$ a Dodgson winner of the election?  That is, is it
  true that for all $d \in C$, $\DodgsonScore(C,c,V) \leq
  \DodgsonScore(C,d,V)$?
\end{quote}

\begin{quote}
  ${\tt Dodgson}$ ${\tt Ranking}$ \\[1ex]
  {\bf Instance:} A preference profile $\pair{C,V}$ and two designated
  candidates $c, d \in C$.\\
  {\bf Question:} Does $c$ tie-or-defeat $d$ in the election?  That is, is it
  true that $\DodgsonScore(C,c,V) \leq \DodgsonScore(C,d,V)$?
\end{quote}

Bartholdi et al.~\cite{bar-tov-tri:j:who-won} established an NP-hardness lower
bound for both these problems.  Their result was optimally improved by
Hemaspaandra, Hemaspaandra, and Rothe~\cite{hem-hem-rot:j:dodgson} who proved
that ${\tt Dodgson}$ ${\tt Winner}$ and ${\tt Dodgson}$ ${\tt Ranking}$ are
complete for~$\parallelnp$, the class of problems solvable in polynomial time
with parallel (i.e., truth-table) access to an NP oracle.

As above, we define the corresponding decision problems for Young elections as
follows.

\begin{quote}
${\tt Young}$ ${\tt Winner}$ \\[1ex]
  {\bf Instance:} A preference profile $\pair{C,V}$ and
  a designated candidate~$c \in C$. \\
  {\bf Question:} Is $c$ a Young winner of the election?  That is, is it
  true that for all $d \in C$, $\YoungScore(C,c,V) \geq
  \YoungScore(C,d,V)$?
\end{quote}

\begin{quote}
${\tt Young}$ ${\tt Ranking}$ \\[1ex]
  {\bf Instance:} A preference profile $\pair{C,V}$ and two designated
  candidates $c, d \in C$.\\
  {\bf Question:} Does $c$ tie-or-defeat $d$ in the election?
  That is, is it true that $\YoungScore(C,c,V) \geq
  \YoungScore(C,d,V)$?
\end{quote}

\subsection{Hardness of Determining Young Winners}
\label{sec:young}

The main result in this section is that the problems ${\tt Young}$ ${\tt
  Winner}$ and ${\tt Young}$ ${\tt Ranking}$ are complete for~$\parallelnp$.
In Theorem~\ref{thm:young-ranking} below, we give a reduction from the problem
${\tt Maximum}$ ${\tt Set}$ ${\tt Packing}$ ${\tt Compare}$ that is defined
below.  For a given familiy $\mathcal{S}$ of sets, let $\kappa(\mathcal{S})$
be the maximum number of pairwise disjoint sets in~$\mathcal{S}$.

\begin{quote}
  ${\tt Maximum}$ ${\tt Set}$ ${\tt Packing}$ ${\tt Compare}$ \\[1ex]
  {\bf Instance:} Two families $\mathcal{S}_1$ and $\mathcal{S}_2$ of sets
  such that, for $i \in \{1,2\}$, each set $S \in \mathcal{S}_i$ is a nonempty
  subset of a given set~$B_i$.\\
  {\bf Question:} Does it hold that $\kappa(\mathcal{S}_1) \geq
  \kappa(\mathcal{S}_2)$?
\end{quote}

To prove that ${\tt Maximum}$ ${\tt Set}$ ${\tt Packing}$ ${\tt Compare}$ is 
$\parallelnp$-complete, we give a reduction from the problem
${\tt Independence}$ ${\tt Number}$ ${\tt Compare}$, 
which has also been used in~\cite{hem-rot-spa:t:vcgreedy}.
Let $G$ be an undirected, simple graph.  An {\em independent set of~$G$\/} is
any subset $I$ of the vertex set of $G$ such that no two vertices in $I$
are adjacent. For any graph~$G$, let $\alpha(G)$ be the 
{\em independence number of~$G$}, i.e., the size of a maximum 
independent set of~$G$.

\begin{quote}
${\tt Independence}$ ${\tt Number}$ ${\tt Compare}$ \\[1ex]
{\bf Instance:} Two graphs $G_1$ and  $G_2$. \\
{\bf Question:} Does it hold that $\alpha(G_1) \geq \alpha(G_2)$?
\end{quote}

Using the techniques of Wagner~\cite{wag:j:min-max}, it can be shown that the
problem ${\tt Independence}$ ${\tt Number}$ ${\tt Compare}$ is
$\parallelnp$-complete; see~\cite[Thm.~12]{spa-vog:c:thetatwo} for an explicit
proof of this result.

\begin{lemma} 
\label{lem:ind-set-compare}
{\rm{}(cf.~\cite{wag:j:min-max,spa-vog:c:thetatwo})}\quad
${\tt Independence}$ ${\tt Number}$ ${\tt Compare}$ is $\parallelnp$-complete.
\end{lemma}

\begin{theorem}
  ${\tt Maximum}$ ${\tt Set}$ ${\tt Packing}$ ${\tt Compare}$ is
  $\parallelnp$-complete.
\end{theorem}

\begin{proof}
We give a polynomial-time many-one reduction from the problem 
${\tt Independence}$ ${\tt Number}$ ${\tt Compare}$
to the problem ${\tt Maximum}$ ${\tt Set}$ ${\tt Packing}$ ${\tt Compare}$.
Let $G_1$ and $G_2$ be two given graphs.  For $i\in\{ 1,2\}$, define $B_i$
  to be the set of edges of~$G_i$, and define $\mathcal{S}_i$ so as to contain
  exactly $\| V(G_i)\|$ subsets of $B_i$: For each vertex $v$ of~$G_i$, add to
  $\mathcal{S}_i$ the set of edges incident to~$v$.  Thus, for each $i\in\{
  1,2\}$, we have $\alpha(G_i) = \kappa(\mathcal{S}_i)$, 
  which proves the theorem.
\end{proof}

\begin{theorem}
\label{thm:young-ranking}
${\tt Young}$ ${\tt Ranking}$ is $\parallelnp$-complete.
\end{theorem}

\begin{proof}
  It is easy to see that ${\tt Young}$ ${\tt Ranking}$ and ${\tt Young}$ ${\tt
    Winner}$ are in~$\parallelnp$.  To prove the $\parallelnp$ lower bound, we
  give a polynomial-time many-one reduction from the problem ${\tt Maximum}$
  ${\tt Set}$ ${\tt Packing}$ ${\tt Compare}$.  
  Let $B_1 = \{ x_1, x_2,\ldots , x_{m}\}$ and $B_2 = \{ y_1, y_2,\ldots ,
  y_{n}\}$ be two given sets, and let $\mathcal{S}_1$ and $\mathcal{S}_2$ be
  given families of subsets of $B_1$ and~$B_2$, respectively.  Recall that
  $\kappa(\mathcal{S}_i)$, for $i\in\{ 1,2\}$, is the maximum number of
  pairwise disjoint sets in~$\mathcal{S}_i$; w.l.o.g., we may assume that
  $\kappa(\mathcal{S}_i) > 2$.

  We define a preference profile $\pair{C,V}$ such that $c$ and $d$ are
  designated candidates in~$C$, and it holds that:
\begin{eqnarray}
\label{eq::kappa1}
\YoungScore(C,c,V) & = & 2 \cdot \kappa(\mathcal{S}_1) + 1 ; \\ 
\label{eq::kappa2}
\YoungScore(C,d,V) & = & 2 \cdot \kappa(\mathcal{S}_2) + 1 .
\end{eqnarray}
Define the set $C$ of candidates as follows:
\begin{itemize}
\item create the two designated candidates $c$ and~$d$;
\item for each element $x_i$ of~$B_1$, create a candidate~$x_i$;
\item for each element $y_i$ of~$B_2$, create a candidate~$y_i$;
\item create two auxiliary candidates, $a$ and~$b$.
\end{itemize}
Define the set $V$ of voters as follows:
\begin{itemize}
\item {\bf Voters representing~\boldmath{$\mathcal{S}_1$}:}\quad For each set
  $E \in \mathcal{S}_1$, create a single voter $v_E$ as follows:
\begin{itemize}
\item Enumerate $E$ as $\{e_1, e_2, \ldots , e_{\|E\|}\}$ (renaming the
  candidates $e_i \in \{ x_1, x_2,\ldots , x_{m}\}$ for notational
  convenience), and enumerate its complement $\overline{E} = B_1 - E$ as
  $\{\overline{e}_1, \overline{e}_2, \ldots , \overline{e}_{m - \|E\|}\}$.

\item To make the preference orders easier to parse, we use
\begin{eqnarray*}
\mbox{``}\overrightarrow{E}\mbox{''} & 
\mbox{to represent the text string} & 
\mbox{``}e_1 > e_2 > \cdots > e_{\|E\|}\mbox{''}; \\
\mbox{``}\overrightarrow{\overline{E}}\mbox{''} & 
\mbox{to represent the text string} & 
\mbox{``}\overline{e}_1 > \overline{e}_2 > \cdots > 
\overline{e}_{m - \|E\|}\mbox{''}; \\
\mbox{``}\overrightarrow{B_1}\mbox{''} & 
\mbox{to represent the text string} & 
\mbox{``}x_1 > x_2 > \cdots > x_m\mbox{''}; \\
\mbox{``}\overrightarrow{B_2}\mbox{''} & 
\mbox{to represent the text string} & 
\mbox{``}y_1 > y_2 > \cdots > y_n\mbox{''}.
\end{eqnarray*}

\item Create one voter $v_E$ with preference order:
\begin{equation}  
\label{eq::1}
\overrightarrow{E} > a > c > \overrightarrow{\overline{E}} > 
\overrightarrow{B_2} > b > d.
\end{equation}
\end{itemize}

\item Additionally, create two voters with preference order:
\begin{equation}
\label{eq::2}
c > \overrightarrow{B_1} > a > \overrightarrow{B_2} > b > d,
\end{equation}
and create $\|S_1\| - 1$ voters with preference order:
\begin{equation}
\label{eq::3}
\overrightarrow{B_1} > c > a >\overrightarrow{B_2} > b  > d.
\end{equation}

\item {\bf Voters representing~\boldmath{$\mathcal{S}_2$}:}\quad For each set
  $F \in \mathcal{S}_2$, create a single voter $v_F$ as follows:
\begin{itemize}
\item Enumerate $F$ as $\{f_1, f_2, \ldots , f_{\|F\|}\}$
  (renaming the candidates $f_j \in \{ y_1, y_2,\ldots , y_{n}\}$ for
  notational convenience), and enumerate its complement $\overline{F} = B_1 -
  F$ as $\{\overline{f}_1, \overline{f}_2, \ldots , \overline{f}_{n -
    \|F\|}\}$.

\item To make the preference orders easier to parse, we use
\begin{eqnarray*}
\mbox{``}\overrightarrow{F}\mbox{''} & 
\mbox{to represent the text string} & 
\mbox{``}f_1 > f_2 > \cdots > f_{\|F\|}\mbox{''}; \\
\mbox{``}\overrightarrow{\overline{F}}\mbox{''} & 
\mbox{to represent the text string} & 
\mbox{``}\overline{f}_1 > \overline{f}_2 > \cdots > 
\overline{f}_{n - \|F\|}\mbox{''}.
\end{eqnarray*}

\item Create one voter $v_F$ with preference order:
\begin{equation}
\label{eq::4}
\overrightarrow{F} > b > d > \overrightarrow{\overline{F}} > 
\overrightarrow{B_1} > a > c.
\end{equation}
\end{itemize}

\item Additionally, create two voters with preference order:
\begin{equation}
\label{eq::5}
d > \overrightarrow{B_2} > b > \overrightarrow{B_1} > a > c,
\end{equation}
and create $\|S_2\| - 1$ voters with preference order:
\begin{equation}
\label{eq::6}
\overrightarrow{B_2} > d > b >\overrightarrow{B_1} > a  > c.
\end{equation}
\end{itemize}

We now prove Equation~(\ref{eq::kappa1}): $\YoungScore(C,c,V) =
2 \cdot \kappa(\mathcal{S}_1) + 1$.

Let $E_1, E_2, \ldots , E_{\kappa(\mathcal{S}_1)} \in \mathcal{S}_1$ be
$\kappa(\mathcal{S}_1)$ disjoint subsets of~$B_1$.  Consider the following
subset  $\widehat{V} \seq V$ of the voters.  $\widehat{V}$ consists of:
\begin{itemize}
\item every voter $v_{E_i}$ corresponding to the set~$E_i$, where $1 \leq i
  \leq \kappa(\mathcal{S}_1)$;
\item the two voters given in Equation~(\ref{eq::2});
\item $\kappa(\mathcal{S}_1) - 1$ voters of the form given in
  Equation~(\ref{eq::3}).
\end{itemize}

Then, $\|\widehat{V}\| = 2 \cdot \kappa(\mathcal{S}_1)+1$.  Note that a strict
majority of the voters in $\widehat{V}$ prefer $c$ over any other candidate,
and thus $c$ is a Condorcet winner in~$\pair{C,\widehat{V}}$.  Hence,
\[
\YoungScore(C,c,V) \geq 2 \cdot \kappa(\mathcal{S}_1) + 1 .
\]

Conversely, to prove that $\YoungScore(C,c,V) \leq 2 \cdot
\kappa(\mathcal{S}_1) + 1$, we need the following lemma.
 
\begin{lemma}
\label{lem:young-lower-bound}
For any $\lambda$ with $3 < \lambda \leq \|S_1\| + 1$, let $V_{\lambda}$ be
any subset of $V$ such that $V_{\lambda}$ contains exactly $\lambda$ voters of
the form~(\ref{eq::2}) or~(\ref{eq::3}) and $c$ is the Condorcet winner
in~$\pair{C, V_{\lambda}}$.  Then, $V_{\lambda}$ contains exactly $\lambda -
1$ voters of the form~(\ref{eq::1}) and no voters of the form~(\ref{eq::4}),
(\ref{eq::5}), or~(\ref{eq::6}).  Moreover, the $\lambda - 1$ voters of the
form~(\ref{eq::1}) in $V_{\lambda}$ represent pairwise disjoint sets
from~$\mathcal{S}_1$.
\end{lemma}

\begin{proofoflemma}{\ref{lem:young-lower-bound}}
  Let $V_{\lambda}$ for fixed $\lambda$ be given as above.  Consider the
  subset of $V_{\lambda}$ that consists of the $\lambda$ voters of the
  form~(\ref{eq::2}) or~(\ref{eq::3}).  Every candidate $x_i$, $1 \leq i \leq
  m$, is preferred to~$c$ by the at least $\lambda - 2$ voters of the
  form~(\ref{eq::3}).  Since $c$ is the Condorcet winner
  in~$\pair{C,V_{\lambda}}$, there exist at least $\lambda - 1 > 2$ voters in
  $V_{\lambda}$ who prefer $c$ to each~$x_i$.  By construction, these voters
  must be of the form~(\ref{eq::1}) or~(\ref{eq::2}).  Since there are at most
  two voters of the form~(\ref{eq::2}), there exists at least one voter of the
  form~(\ref{eq::1}), say~$\tilde{v}$.  Since the voters of the
  form~(\ref{eq::1}) represent~$\mathcal{S}_1$, which does not contain empty
  sets, there exists some candidate $x_j$ who is preferred to $c$
  by~$\tilde{v}$.  In particular, $c$ must outpoll $x_j$ in
  $\pair{C,V_{\lambda}}$ and thus needs more than $(\lambda - 2) + 1$ votes of
  the form~(\ref{eq::1}) or~(\ref{eq::2}).  There are at most two voters of
  the form~(\ref{eq::2}); hence, $c$ must be preferred by at least $\lambda -
  2$ voters of the form~(\ref{eq::1}) that are distinct from~$\tilde{v}$.
  Summing up, $V_{\lambda}$ contains at least $\lambda - 1$ voters of the
  form~(\ref{eq::1}).
  
  On the other hand, since $c$ is the Condorcet winner in~$\pair{C,
    V_{\lambda}}$, $c$ must in particular outpoll~$a$, who is preferred to $c$
  by the at most $\lambda$ voters of the form~(\ref{eq::2}) or~(\ref{eq::3}).
  Hence, $V_{\lambda}$ may contain at most $\lambda - 1$ voters of the
  form~(\ref{eq::1}), (\ref{eq::4}), (\ref{eq::5}), or~(\ref{eq::6}).  It
  follows that $V_{\lambda}$ contains exactly $\lambda - 1$ voters of the
  form~(\ref{eq::1}) and no voters of the form~(\ref{eq::4}), (\ref{eq::5}),
  or~(\ref{eq::6}).
  
  For a contradiction, suppose that there is a candidate $x_j$ who is
  preferred to $c$ by more than one voter of the form~(\ref{eq::1})
  in~$V_{\lambda}$.  Then, $c$ is preferred to $x_j$ by at most two voters of
  the form~(\ref{eq::2}) and by at most $(\lambda - 1) - 2 = \lambda - 3$
  voters of the form~(\ref{eq::1}); $x_j$ is preferred to $c$ by at least
  $\lambda - 2$ voters of the form~(\ref{eq::3}) and by at least two 
  voters of the form~(\ref{eq::1}).  Since $c$ thus has at most $\lambda - 1$
  votes and $x_j$ has at least $\lambda$ votes in~$V_{\lambda}$, $c$ is not a
  Condorcet winner in $\pair{C, V_{\lambda}}$, a contradiction.  Thus, every
  candidate~$x_i$, $1 \leq i \leq m$, is preferred to $c$ by at most one voter
  of the form~(\ref{eq::1}) in~$V_{\lambda}$, which means that the $\lambda -
  1$ voters of the form~(\ref{eq::1}) in $V_{\lambda}$ represent pairwise
  disjoint sets from~$\mathcal{S}_1$.~\qedoflemma{\ref{lem:young-lower-bound}}
\end{proofoflemma}

To continue the proof of Theorem~\ref{thm:young-ranking}, let $k =
\YoungScore(C,c,V)$.  Let $\widehat{V} \seq V$ be a subset of size $k$ such
that $c$ is the Condorcet winner in~$\pair{C,\widehat{V}}$.  Suppose that
there are exactly $\lambda \leq \|S_1\| + 1$ voters of the form~(\ref{eq::2})
or~(\ref{eq::3}) in~$\widehat{V}$.  Since~$c$, the Condorcet winner
of~$\pair{C,\widehat{V}}$, must in particular outpoll~$a$, we have $\lambda
\geq \left\lceil\frac{k+1}{2}\right\rceil$.  By our assumption that
$\kappa(\mathcal{S}_1) > 2$, it follows from $k \geq 2 \cdot
\kappa(\mathcal{S}_1) + 1$ that $\lambda > 3$.
Lemma~\ref{lem:young-lower-bound} then implies that there are exactly $\lambda
- 1$ voters of the form~(\ref{eq::1}) in~$\widehat{V}$ that represent pairwise
disjoint sets from~$\mathcal{S}_1$, and $\widehat{V}$ contains no voters of
the form~(\ref{eq::4}), (\ref{eq::5}), or~(\ref{eq::6}).  Hence, $k = 2 \cdot
\lambda - 1$ is odd, and $\frac{k-1}{2} = \lambda - 1 \leq
\kappa(\mathcal{S}_1)$, which proves Equation~(\ref{eq::kappa1}).

Equation~(\ref{eq::kappa2}) can be proven analogously.  Thus, we have
$\kappa(\mathcal{S}_1) \geq \kappa(\mathcal{S}_2)$ if and only if
$\YoungScore(C,c,V) \geq \YoungScore(C,d,V)$, which completes the proof of
Theorem~\ref{thm:young-ranking}.
\end{proof}

\begin{theorem}  
\label{thm:young-winner}
${\tt Young}$ ${\tt Winner}$ is $\parallelnp$-complete.
\end{theorem}
\begin{proof}
  To prove the theorem, we modify the reduction from
  Theorem~\ref{thm:young-ranking} to a reduction from the problem ${\tt
    Maximum}$ ${\tt Set}$ ${\tt Packing}$ ${\tt Compare}$ to the problem ${\tt
    Young}$ ${\tt Winner}$ as follows.  Let $\pair{C,V}$ be the preference
  profile constructed in the proof of Theorem~\ref{thm:young-ranking} with the
  designated candidates $c$ and~$d$.  We alter this profile such that all
  other candidates do worse than $c$ and~$d$.
  
  From $\pair{C,V}$, we construct a new preference profile~$\pair{D,W}$.  To
  define the new set $D$ of candidates, 
  replace every candidate $g\in C$ except $c$ and $d$ by
  $\|V\|$ candidates $g^1, g^2, \ldots , g^{\|V\|}$.
   
  To define the new voter set~$W$, replace each occurrence of candidate~$g$ in
  the $i$-th preference order of $V$ by the text string:
\begin{displaymath}
g^{i\mod \| V\|} > g^{i + 1\mod \| V\|} > g^{i + 2\mod \| V\|} > 
\cdots  > g^{i + \| V\| - 1 \mod \| V\|} .
\end{displaymath}
It is easy to see that this modification does not change the Young score of
$c$ and~$d$.  On the other hand, the Young score of any other candidate now
is at most~$1$.
Thus, there is no candidate $h$ with $\YoungScore(C,h,V) >
\YoungScore(C,c,V)$ or $\YoungScore(C,h,V) > \YoungScore(C,d,V)$.  Hence,
$\kappa(\mathcal{S}_1) \geq \kappa(\mathcal{S}_2)$ if and only if $c$ is a
winner of the election~$\pair{D,W}$.
\end{proof}

\section{Homogeneous Young and Dodgson Voting Schemes}
\label{sec:homogeneous}

Social choice theorists have studied many ``reasonable'' properties that any
``fair'' election procedure arguably should satisfy, including very natural
properties such as nondictatorship, monotonicity, the Pareto Principle, and
independence of irrelevant alternatives.  One of the most notable results in
this regard is Arrow's famous Impossibility
Theorem~\cite{arr:b:polsci:social-choice} stating that the just-mentioned four
properties are logically inconsistent, and thus no ``fair'' voting scheme can
exist.

In this section, we are concerned with another quite natural property, the
homogeneity of voting schemes
(see~\cite{fis:j:condorcet,you:j:extending-condorcet}).  

\begin{definition}
A voting scheme $f$
is said to be {\em homogeneous\/} if and only if for each preference
profile~$\pair{C,V}$ and for all positive integers~$q$, it holds that
$f(\pair{C,V}) = f(\pair{C,qV})$, where $qV$ denotes $V$ replicated $q$ times.
\end{definition}

That is, homogeneity means that splitting each voter $v \in V$ into $q$ 
voters, each of whom has the same preference order as~$v$, yields exactly the
same choice set of winning candidates.

Fishburn~\cite{fis:j:condorcet} showed that neither the Dodgson nor the Young
voting schemes are homogeneous.  For the Dodgson SCF, he presented a
counterexample with seven voters and eight candidates; for the Young SCF, he
modified a preference profile constructed by 
Young with 37 voters and five candidates.
Fishburn~\cite{fis:j:condorcet} provided the following limit devise in order
to define homogeneous variants of the Dodgson and Young SCFs.  For example,
the Dodgson scheme can be made homogeneous by defining from the function
$\DodgsonScore$ for each preference profile $\pair{C,V}$ and designated
candidate~$c \in C$ the function
\[
   \DodgsonScore^*(C,c,V) = \lim_{q \rightarrow \infty} 
   \frac{\DodgsonScore(C,c,q V)}{q} .
\]
The resulting SCF is denoted by Dodgson$^*$ SCF, and the corresponding winner and
ranking problems are denoted by ${\tt Dodgson}^*$ ${\tt Winner}$ and ${\tt
  Dodgson}^*$ ${\tt Ranking}$.  Analogously, the Young voting scheme defined
in Section~\ref{sec:issues} can be made homogeneous by defining
$\YoungScore^*$.  Remarkably, Young~\cite{you:j:extending-condorcet} showed
that the corresponding problem ${\tt Young}^*$ ${\tt Winner}$ can be solved by
a linear program.  Hence, the problem ${\tt Young}^*$ ${\tt Winner}$ is
efficiently solvable, since the problem ${\tt Linear}$ ${\tt Programming}$ can
be decided in polynomial time~\cite{hac:j:linear-programming}, see
also~\cite{kar:j:linear-programming}.  We establish an analogous result for
the problems ${\tt Dodgson}^*$ ${\tt Winner}$ and ${\tt Dodgson}^*$ ${\tt
  Ranking}$ below.

\begin{theorem}
\label{sec:}
${\tt Dodgson}^*$ ${\tt Winner}$ and ${\tt Dodgson}^*$ ${\tt Ranking}$ can be
solved in polynomial time.
\end{theorem}

\begin{proof}
  Bartholdi, Tovey, and Trick~\cite{bar-tov-tri:j:who-won} provided an integer
  linear program for determining the Dodgson score of a given candidate~$c$.
  They noted that if the number of candidates is fixed, then the winner
  problem for Dodgson elections (in the inhomogeneous case defined in
  Section~\ref{sec:issues}) can be solved in polynomial time using the
  algorithm of Lenstra~\cite{len:j:integer-linear-programming}.
  
  Based on their integer linear program, we provide a linear program for
  computing $\DodgsonScore^*(C,c,V)$ for a given preference profile
  $\pair{C,V}$ and a given candidate~$c$.  Since ${\tt Linear}$ ${\tt
    Programming}$ is polynomial-time solvable~\cite{hac:j:linear-programming},
  it follows that the problems ${\tt Dodgson}^*$ ${\tt Winner}$ and ${\tt
    Dodgson}^*$ ${\tt Ranking}$ can be solved in polynomial time, even if the
  number of candidates is not prespecified.
  
  Let a profile $\pair{C,V}$ and a candidate~$c \in C$ be given, and let $V =
  \{v_1 , v_2 , \ldots , v_n\}$.  Our linear program has the variables $x_{i,
    j}$, $e_{i,j,k}$, and~$w_k$, where $1 \leq i \leq n$, $1 \leq j \leq \|C\|
  - 1$, and $k \in C - \{c\}$.  Then, $\DodgsonScore^*(C,c,V)$ is the value of
  the linear program
\begin{eqnarray}
\label{eq:linear-program}
  \min \sum_{i,j} j \cdot x_{i,j} 
\end{eqnarray}
subject to the constraints:
\begin{itemize}
\item[(1)] $\sum_{j} x_{i,j} = 1$ for each voter~$v_i$;
\item[(2)] $\sum_{i,j} e_{i,j,k} \cdot x_{i,j} + w_k > \frac{n}{2}$ for each
  candidate $k \in C - \{c\}$;
\item[(3)] $0 \leq x_{i,j} \leq 1$ for each $i$ and~$j$.
\end{itemize}
The variables and constraints can be interpreted as follows.  For given $i$
and~$j$, $x_{i,j}$ is a rational number in the interval $[0, 1]$ that gives
the percentage~$\frac{v_{i,j}^{q}}{q}$, where $q$ is the least common multiple
of the denominators in all~$x_{i,j}$, and $v_{i,j}^{q}$ is the number of
voters among the $q$ replicants of voter $v_i$ in which $c$ is moved upwards
by $j$ positions.  For given $i$, $j$, and~$k$, $e_{i,j,k} = 1$ if the result
of moving $c$ upwards by $j$ positions in the preference order of voter $v_i$
is that $c$ gains one additional vote against candidate~$k$, and $e_{i,j,k} =
0$ otherwise.  For any candidate $k$ other than~$c$, $w_k$ gives the number of
voters who prefer $c$ over~$k$.  Hence, the set of constraints~(2) ensures
that $c$ becomes a Condorcet winner.  The set of constraints~(1) ensures that
$v_{i,j}^{q}$, summed over all possible positions~$j$, equals the number $q$ of
all replicants of voter~$v_i$.  
The objective is to minimize the number of switches needed to make $c$ a
Condorcet winner.  For the homogeneous case of Dodgson elections, the linear
program~(\ref{eq:linear-program}) tells us how many times we have to replicate
each voter $v_i$ (namely, $q$ times) and in how many of the replicants of each
voter~$v_i$ the given candidate $c$ has to be moved upwards by how many
positions in order to achieve this objective.
\end{proof}

\bibliographystyle{alpha}

\bibliography{/home/inf1/rothe/BIGBIB/joergbib}

\end{document}